\documentclass[12pt,a4paper]{article}

\usepackage{amsmath,amssymb,bm}
\usepackage{graphicx,color}
\usepackage{cite}
\usepackage[compat=1.1.0]{tikz-feynhand}
\usepackage[hang,small,bf]{caption}
\usepackage[subrefformat=parens]{subcaption}

\allowdisplaybreaks

\usepackage[height=8.85in,width=6.4in]{geometry}

\newcommand\dd{\mathrm{d}}

\definecolor{BlueViolet}{rgb}{0.2, 0.00, 0.7}
\definecolor{Blue}{rgb}{0.15, 0.00, 0.9}
\usepackage[setpagesize=false,bookmarksnumbered=true,bookmarksopen=true,colorlinks=true,linkcolor=Blue,citecolor=Blue,urlcolor=BlueViolet]{hyperref}

\begin{document}
\begin{titlepage}
\setcounter{page}{0} 

\begin{center}

\vskip .55in

\begingroup
\centering

\hfill {\tt STUPP-23-263}
\vskip .55in

{\large\bf Contribution of Majoron to Hubble tension\\
in gauged U(1)$_{L_\mu-L_\tau}$ Model}

\endgroup

\vskip .4in

\renewcommand{\thefootnote}{\fnsymbol{footnote}}
{
Kento Asai$^{(a)}$\footnote{
  \href{mailto:kento@icrr.u-tokyo.ac.jp}
  {\tt kento@icrr.u-tokyo.ac.jp}},
Tomoya Asano$^{(b)}$\footnote{
  \href{mailto:asano@krishna.th.phy.saitama-u.ac.jp}
  {\tt asano@krishna.th.phy.saitama-u.ac.jp}},
Joe Sato$^{(c)}$\footnote{
  \href{mailto:sato-joe-mc@ynu.ac.jp}
  {\tt sato-joe-mc@ynu.ac.jp}}, and
Masaki J. S. Yang$^{(b)}$\footnote{
  \href{mailto:mjsyang@mail.saitama-u.ac.jp}
  {\tt mjsyang@mail.saitama-u.ac.jp}}
}

\vskip 0.2in

\begingroup\small
\begin{minipage}[t]{0.9\textwidth}
\centering\renewcommand{\arraystretch}{0.9}
{\it
\begin{tabular}{c@{\,}l}
$^{(a)}$
& Institute for Cosmic Ray Research (ICRR), The University of Tokyo, Kashiwa,\\
& Chiba 277--8582, Japan \\[2mm]
$^{(b)}$
& Department of Physics, Faculty of Science, Saitama University, Saitama 338--8570, \\ 
& Japan \\ [2mm]
$^{(c)}$
& Department of Physics, Faculty of Engineering Science, Yokohama National University, \\
& Yokohama 240--8501, Japan \\
\end{tabular}
}
\end{minipage}
\endgroup

\end{center}

\vskip .4in

\begin{abstract}\noindent
In this paper, we analyze parameter regions that can alleviate the Hubble tension in the U(1)$_{L_\mu - L_\tau}$ model with the broken lepton number U(1)$_L$ symmetry.
As new particles, this model has a U(1)$_{L_\mu - L_\tau}$ gauge boson $Z'$ and a Majoron $\phi$, which can affect the early universe and the effective number of neutrino species $N_{\rm eff}$. 
If $Z'$ and $\phi$ simultaneously exist in the early universe, $Z'\,$--$\,\phi$ interaction processes such as $Z'\nu_\alpha \leftrightarrow \phi\bar{\nu}_\beta$ occur. 
The comparison of $N_{\rm eff}$ between the cases with and without the $Z'\,$--$\,\phi$ interaction processes shows that these processes make a small contribution of $\mathcal{O}(10^{-4})$ to $N_{\rm eff}$, and it does not need to be considered for the alleviation of the Hubble tension.
Based on these facts, we calculated $N_{\rm eff}$ for various Majoron parameters without the $Z'\,$--$\,\phi$ interaction processes to search parameters that could alleviate the Hubble tension. 
As a result, we found that the U(1)$_{L_\mu - L_\tau}$ gauge boson and Majoron can alleviate the Hubble tension in some parameter regions, and there is a non-trivial synergy contribution between $Z'$ and $\phi$.
Moreover, the parameter region with a lighter mass $m_\phi \lesssim 2$\,MeV and a larger coupling $\lambda \gtrsim 10^{-8}$ is excluded because it predicts too large $N_{\rm eff}$, i.e. $N_{\rm eff} \gtrsim 3.5$.
The favored and restricted regions of the Majoron parameters depend on the $Z'$ parameters because of the presence of the $Z'$ contribution and synergy one.

\end{abstract}
\end{titlepage}
\setcounter{page}{1}
\renewcommand{\thefootnote}{\#\arabic{footnote}}
\setcounter{footnote}{0}


\section{Introduction}
\label{sec:introduction}

The Hubble constant $H_0$ has been measured in the last decades from two different approaches.
One is the measurements of the local universe. 
The Hubble constant is directly obtained by measuring distances and velocities of celestial bodies~\cite{Riess:2018uxu,Riess:2018byc,Riess:2019cxk,Riess:2021jrx}, and SH0ES program reported $H_0 = 73.04 \pm 1.04\,\textrm{km/s/Mpc}$~\cite{Riess:2021jrx}.
The other is the analyses of the cosmic microwave background (CMB) under the assumption of the $\Lambda$CDM cosmology, and the Planck collaboration reported $H_0 = 67.36 \pm 0.54\,\textrm{km/s/Mpc}$~\cite{Planck:2018vyg}.
The discrepancy between two approaches reaches the level of around $4\sigma - 6\sigma$ and is well-known as the Hubble tension.

Although this discrepancy may be originated from systematic errors in the measurements of the local Universe~\cite{Efstathiou:2013via,Freedman:2017yms,Ivanov:2020mfr}, various modifications of the $\Lambda$CDM cosmology based on physics beyond the Standard Model (SM) have been proposed.
It is well-known as a simple way to relax this tension is to introduce new contributions to the radiation component of the energy density of the universe and increase the effective number of neutrino species $N_{\rm eff}$~\cite{Shakya:2016oxf,Berlin:2018ztp,DEramo:2018vss,Dessert:2018khu,Escudero:2019gzq,Hooper:2019gtx,Blinov:2019gcj,Gelmini:2019deq,Escudero:2019gvw}.\footnote{
It should be mentioned that increasing $N_{\rm eff}$ worsens another milder tension relative to $\sigma_8$~\cite{Planck:2018vyg,Hildebrandt:2018yau} which is the cosmological parameter about the matter density fluctuation amplitude on 8\,Mpc scales.
}
Among them, the alleviation based on the gauged U(1)$_{L_\mu-L_\tau}$ model~\cite{Escudero:2019gzq} is also implied by the muon anomalous magnetic moment (muon $g-2$). 
The measurements of the muon $g-2$ at Fermilab~\cite{Muong-2:2021ojo,Muong-2:2021vma} and the Brookhaven National Laboratory~\cite{Muong-2:2006rrc} have reported the experimental averaged value which has a discrepancy from the theoretical calculations of the SM prediction~\cite{Aoyama:2020ynm} as follows~:
\begin{align}
    \Delta a_\mu 
    = a_\mu^{\rm exp} - a_\mu^{\rm SM}
    = 251(59) \times 10^{-11}~,
\end{align}
whose significance is about to $4.2\sigma$.
Additionally, the planned experiment at J-PARC, which will use an alternative approach and ultra-cold muons~\cite{Mibe:2011zz}, will offer independent data about the systematic uncertainties.

In the gauged U(1)$_{L_\mu-L_\tau}$ models~\cite{Foot:1990mn,He:1990pn,He:1991qd,Foot:1994vd}, the difference between the $\mu$ and $\tau$ flavor numbers is gauged.
The U(1)$_{L_\mu-L_\tau}$ gauge boson couples to the muon, but does not to the electron and quarks at the tree level.
Therefore, the U(1)$_{L_\mu-L_\tau}$ gauge boson can explain the discrepancy of the muon $g-2$, evading the severe experimental constraints~\cite{Ibe:2016dir,Kaneta:2016uyt,Araki:2017wyg}.
However, it is known that the minimal extended model by the U(1)$_{L_\mu-L_\tau}$ gauge symmetry has a strong correlation between the neutrino oscillation parameters, such as the mixing angles and mass squared differences, and the CP phases and sum of the neutrino masses~\cite{Asai:2017ryy,Asai:2018ocx,Asai:2019ciz}.
There is a tension between the global fit of the neutrino oscillation parameters~\cite{Esteban:2018azc,nufit} and the bound on the sum of the neutrino masses by the Planck observation~\cite{Planck:2018vyg,RoyChoudhury:2019hls,Ivanov:2019hqk}.
Therefore, non-minimal gauged U(1)$_{L_\mu-L_\tau}$ models with multiple scalars are considered (for instance, see Ref.~\cite{Araki:2019rmw}) and give a rich phenomenology.

There is a possibility that such non-minimal gauged U(1)$_{L_\mu-L_\tau}$ models also have a global lepton number symmetry U(1)$_L$ as the model in Ref.~\cite{Araki:2019rmw} does.
When U(1)$_L$ is softly broken, models have not only a U(1)$_{L_\mu-L_\tau}$ gauge boson but also a pseudo-Nambu-Goldstone boson (pNGB) called Majoron.
The Majoron with keV mass can also contribute to the alleviation of the Hubble tension~\cite{Escudero:2019gvw}, and the contribution of the U(1)$_{L_\mu-L_\tau}$ gauge boson and Majoron to the effective neutrino species and the relaxation of the Hubble tension have been discussed in Ref.~\cite{Araki:2021xdk}.
However, it has been assumed in Ref.~\cite{Araki:2021xdk} that Majorons are produced after the U(1)$_{L_\mu-L_\tau}$ gauge bosons decay, that is, the U(1)$_{L_\mu-L_\tau}$ gauge boson and Majoron do not exist simultaneously.
In this paper, we consider the parameter region where the U(1)$_{L_\mu-L_\tau}$ gauge boson and Majoron exist in the same period and reanalyze the possibility of the alleviation of the Hubble tension in the gauged U(1)$_{L_\mu-L_\tau}$ model with a Majoron based on Ref.~\cite{Araki:2019rmw}.

The rest of the paper is organized as follows. 
In Section~\ref{sec:model}, we introduce the gauged U(1)$_{L_\mu-L_\tau}$ model with the global U(1)$_L$ symmetry.
In Section~\ref{sec:evolution}, we explain how to obtain the time evolution of temperatures in the early universe.
In Section~\ref{sec:rate}, we describe the newly incorporated contribution, $Z'$\,--\,$\phi$ scattering.
In Section~\ref{sec:result}, we discuss the contribution of $Z'$\,--\,$\phi$ scattering, and analyze the parameter region of Majoron that can alleviate Hubble tension.
Finally, we conclude in Section~\ref{sec:summary}.

\section{\texorpdfstring{U(1)$_{L_\mu-L_\tau} \times$ U(1)$_L$ model}{U(1)Lμ-Lτ × U(1)L model}}
\label{sec:model}

In this paper, we consider the extension of the SM by the U(1)$_{L_\mu-L_\tau}$ gauge symmetry and the global lepton number symmetry U(1)$_L$, and this model has a U(1)$_{L_\mu-L_\tau}$ gauge boson and a Majoron.
The Lagrangian relevant to the U(1)$_{L_\mu-L_\tau}$ gauge boson $Z'^\rho$ and Majoron $\phi$ is given by
\begin{align}
   \mathcal{L} &= \mathcal{L}_{\rm SM} + \mathcal{L}_{Z'} + \mathcal{L}_\phi~, \\
\label{eq:Lag-Zmt}
    \mathcal{L}_{Z'}
    &= -\frac{1}{4} Z'^{\rho\sigma} Z'_{\rho\sigma} + \frac{1}{2} m_{Z'}^2 Z'^\rho Z'_\rho + g_{Z'}^{} Z'^\rho J_{Z'}^\rho + \epsilon e {Z'}_\mu J_{\rm EM}^\mu~, \\
\label{eq:Lag-phi}
    \mathcal{L}_\phi 
    &= -\frac{1}{2}m _\phi^2 \phi^2 + (h_{\alpha\beta} \bar{\nu}_\alpha \nu^c_\beta \phi + \mathrm{H.c.})~, 
\end{align}
where $Z'^{\rho\sigma}$ denotes the field strength tensor of the U(1)$_{L_\mu-L_\tau}$ gauge boson, $m_{Z'}^{}\,(m_\phi)$ the mass of $Z'\,(\phi)$, and $\nu_\alpha^c \equiv (\nu_\alpha^{})^c = C \bar{\nu}_\alpha^T$ with $C$ being the charge conjugation matrix.
In Eq.~\eqref{eq:Lag-Zmt}, the electromagnetic and U(1)$_{L_\mu-L_\tau}$ currents are written by
\begin{align}
\label{eq:EM-current}
    J_{\rm EM}^\rho &= 
    \sum_{i=1,2,3} \left( \frac{2}{3} \bar{u}_i \gamma^\rho u_i - \frac{1}{3} \bar{d}_i \gamma^\rho d_i - \bar{e}_i \gamma^\rho e_i \right)~, \\
    J_{Z'}^\rho &= 
    \bar{\mu} \gamma^\rho \mu + \bar{\nu}_\mu \gamma^\rho P_L \nu_\mu - \bar{\tau} \gamma^\rho \tau - \bar{\nu}_\tau \gamma^\rho P_L \nu_\tau~,
\end{align}
where $i$ denotes the index of the fermion generation.

We assume that the kinetic mixing between the electromagnetic and U(1)$_{L_\mu-L_\tau}$ gauge bosons is vanishing at some high scale.
Even for this case, the kinetic mixing radiatively appears, and the one-loop contribution is given by~\cite{Kamada:2015era}
\begin{align}
\label{eq:mixing}
    \epsilon &\simeq 
    \frac{e g_{Z'}^{}}{12\pi^2} \log \frac{m_\tau^2}{m_\mu^2}~.
\end{align}
In this paper, we focus on the the U(1)$_{L_\mu-L_\tau}$ gauge boson with the mass $m_{Z'}^{} < 2 m_\mu$ for the successful explanation of the muon $g-2$, and in this case, the decay width of the U(1)$_{L_\mu-L_\tau}$ gauge boson is given by~\footnote{
In this paper, we assume that the neutrinos are Majorana particles and massless.
}
\begin{align}
\label{eq:Zp-width}
    \Gamma_{Z'} = \Gamma(Z' \to e^+ e^-) + \sum_{\alpha = e, \mu, \tau} \Gamma(Z' \to \nu_\alpha \bar{\nu}_\alpha)~,
\end{align}
where
\begin{align}
\label{eq:Zp2ee}
    \Gamma(Z' \to e^+ e^-) &=
    \frac{\epsilon^2 e^2}{12 \pi} m_{Z'}^{} \left( 1 + \frac{2 m_e^2}{m_{Z'}^2} \right) \sqrt{1 - \frac{4 m_e^2}{m_{Z'}^2}}~, \\
\label{eq:Zp2nunu}
    \Gamma(Z' \to \nu_\alpha \bar{\nu}_\alpha) &=
    \frac{g_{Z'}^2}{24 \pi} m_{Z'}~.
\end{align}

The Majoron decays into two neutrinos or anti-neutrinos, and the decay width is given by
\begin{align}
\label{eq:phi2nunu}
    \Gamma_\phi 
    = \sum_{\alpha, \beta = e, \mu, \tau} \left\{ \Gamma(\phi \to \nu_\alpha \nu_\beta) + \Gamma(\phi \to \bar{\nu}_\alpha \bar{\nu}_\beta) \right\} \nonumber 
    = \frac{\lambda^2}{4 \pi} m_\phi~,
\end{align}
with $\lambda^2 \equiv \operatorname{tr}(h^\dag h)$.

\section{Time evolution of temperatures}
\label{sec:evolution}

In this paper, we focus on the deviation of the effective number of neutrino species in the simultaneous presence of the U(1)$_{L_\mu-L_\tau}$ gauge boson and Majoron.
The effective number of neutrino species in the U(1)$_{L_\mu-L_\tau} \times$ U(1)$_L$ model is divided into four components as follows~:
\begin{equation}
\label{eq:Neffs}
    N_{\rm eff} =
    N_{\rm eff}^{\rm SM} + \Delta N_{\rm eff}^{Z'} + \Delta N_{\rm eff}^{\phi} + \Delta N_{\rm eff}^{Z'\phi}~,
\end{equation}
where $N_{\rm eff}^{\rm SM} \simeq 3.045$ is the SM prediction of $N_{\rm eff}$~\cite{deSalas:2016ztq,Akita:2020szl}, and $\Delta N_{\rm eff}^{Z'} (\Delta N_{\rm eff}^{\phi})$ is the deviation of $N_{\rm eff}$ in the presence of only the U(1)$_{L_\mu-L_\tau}$ gauge boson (Majoron) from the value of the SM prediction $N_{\rm eff}^{\rm SM}$.
The last term in Eq.~\eqref{eq:Neffs}, $\Delta N_{\rm eff}^{Z'\phi}$, is the contribution which does not appear until both the U(1)$_{L_\mu-L_\tau}$ gauge boson and Majoron are contained in the models.
The effective number of neutrino species is defined by
\begin{align}
\label{eq:Neff}
    \left. 
    N_{\rm eff} =
    3 \left( \frac{11}{4} \right)^{\frac{4}{3}} \left( \frac{T_\nu}{T_\gamma} \right)^4 \right|_{T_\gamma = T_*}~,
\end{align}
where $T_*$ denotes the temperature at which $T_\gamma$ and $T_\nu$ become constant.
For estimation of the contributions of $Z'$ and $\phi$ to $N_{\rm eff}$, it is necessary to follow the time evolution of the temperatures of the photon and neutrino in the simultaneous presence of the U(1)$_{L_\mu-L_\tau}$ gauge boson and Majoron.
In this section, we explain how to follow the time evolution of the temperatures of the related particles.
For the successful explanation of the muon $g-2$, we fix the gauge boson mass and coupling of the U(1)$_{L_\mu-L_\tau}$ gauge symmetry as the following three parameter sets~: $(m_{Z'}^{}, g_{Z'}^{}) = (13\,{\rm MeV}, 5.0\times 10^{-4})$, $(18\,{\rm MeV}, 4.0 \times 10^{-4})$, and $(100\,{\rm MeV}, 7.0 \times 10^{-4})$.

The contribution of the U(1)$_{L_\mu-L_\tau}$ gauge boson to $N_{\rm eff}$ in the absence of the Majoron has already been discussed in Refs.~\cite{Escudero:2019gzq,Araki:2021xdk} under the following approximations~\footnote{
The detailed discussion on the justification of the assumptions is shown in Refs.~\cite{Escudero:2019gzq,Escudero:2018mvt,EscuderoAbenza:2020cmq}.
}~:
\begin{enumerate}
\item 
The distribution functions of all particles follow the thermal equilibrium ones.
\item
In the estimation of the collision terms, the Maxwell-Boltzmann distribution is used, and the mass of the electron is ignored.
\item
Chemical potentials of all particles, $\mu_i$, are ignored.
\item 
The masses of the active neutrinos are ignored.
\item 
Particles in the same thermal bath have the same temperature, namely, $T_\gamma = T_{e^\pm}$ and $T_{Z'} = T_\nu$ with $T_\nu \equiv T_{\nu_e} = T_{\nu_\mu} = T_{\nu_\tau}$.\footnote{
The time scale of the neutrino oscillation is much faster than that of the interactions between the neutrinos and the other particles.
Therefore, the neutrino oscillation makes the distribution functions of $\nu_e$ and $\nu_{\mu,\tau}$ equilibrate, and the temperatures of $\nu_e$ and $\nu_{\mu,\tau}$ become the same value~\cite{Escudero:2018mvt}.
}
\end{enumerate}
Under these approximations, the following equations of the time evolution are derived from the Boltzmann equations~\cite{Araki:2021xdk}~:
\begin{align}
   \frac{\dd T_\gamma}{\dd t} &=
   - \left( \frac{\partial \rho_\gamma}{\partial T_\gamma} + \frac{\partial \rho_e}{\partial T_\gamma} \right)^{-1} \left[ 4 H \rho_\gamma + 3 H (\rho_e + P_e) + \frac{\delta \rho_\nu}{\delta t} + \frac{\delta \rho_{Z'}^{}}{\delta t} + \frac{\delta \rho_\phi}{\delta t} \right]~,
\label{eq:Boltzmann-gamma} \\
   \frac{\dd T_\nu}{\dd t} &=
   - \left( \frac{\partial \rho_{Z'}^{}}{\partial T_\nu} + \frac{\partial \rho_\nu}{\partial T_\nu} \right)^{-1} \left[ 3 H (\rho_{Z'}^{} + P_{Z'}) + 4 H \rho_\nu - \frac{\delta \rho_{Z'}^{}}{\delta t} - \frac{\delta \rho_\nu}{\delta t} \right]~,
\label{eq:Boltzmann-nu} \\
   \frac{\dd T_\phi}{\dd t} &=
   - \left( \frac{\partial \rho_\phi}{\partial T_\phi} \right)^{-1} \left[ 3 H (\rho_\phi + P_\phi) - \frac{\delta \rho_\phi}{\delta t} \right]~,
\label{eq:Boltzmann-phi}
\end{align}
where $\rho_i$ and $P_i$ are the energy density and pressure of the particle $i$, respectively, and $\delta \rho_i/\delta t$ is called the energy transfer rate.

The energy transfer rates for the neutrino, U(1)$_{L_\mu-L_\tau}$ gauge boson, and Majoron are obtained from the energy conservation law of the following processes~:
\begin{itemize}
\item
Weak interactions in the SM
\begin{equation*}
    \nu_\alpha \bar{\nu}_\alpha \leftrightarrow e^- e^+,~ \nu_\alpha e^\pm \leftrightarrow \nu_\alpha e^\pm,~ \bar{\nu}_\alpha e^\pm \leftrightarrow \bar{\nu}_\alpha e^\pm~,
\end{equation*}
\item Decay and inverse decay of $Z'$
\begin{equation*}
    Z' \leftrightarrow e^- e^+,~ Z' \leftrightarrow \nu_{\alpha'} \bar{\nu}_{\alpha'} \quad (\alpha' = \mu, \tau)~,
\end{equation*}
\item Decay and inverse decay of Majoron
\begin{equation*}
    \phi \leftrightarrow \nu_\alpha \nu_\beta,~ \phi \leftrightarrow \bar{\nu}_\alpha \bar{\nu}_\beta~,
\end{equation*}
\item $Z'$\,--\,$\phi$ interactions
\begin{equation*}
    Z'\nu_{\alpha} \leftrightarrow \phi \bar{\nu}_\beta,~ Z' \bar{\nu}_{\alpha} \leftrightarrow \phi \nu_\beta,~ Z'\phi \leftrightarrow \nu_\alpha \nu_\beta,~ Z' \phi \leftrightarrow \bar{\nu}_\alpha \bar{\nu}_\beta~.
\end{equation*}
\end{itemize}
The explicit expressions of the energy transfer rates other than $Z'\,$--$\,\phi$ interactions are obtained in Ref.~\cite{Araki:2021xdk}. 
Those of $Z'\,$--$\,\phi$ interactions are discussed in the next section and Appendix~\ref{append:integ}.

\section{Energy transfer rates}
\label{sec:rate}

In this section, we show the calculation of the energy transfer rates of $Z'\,$--$\,\phi$ interactions, which occur in the simultaneous presence of the U(1)$_{L_\mu-L_\tau}$ gauge boson and Majoron.
The energy transfer rate is calculated by integration of the collision term with respect to the momentum of the interested particle, $a$, as follows\,:
\begin{align}
\label{eq:Etransfer}
   \frac{\delta \rho_a}{\delta t} =
   \int \frac{\dd \bm{p}_a}{(2\pi)^3} E_a C[f_a(\bm{p}_a)]~.
\end{align}
Here, $E_a$ and $\bm{p}_a$ stand for the energy and three-dimensional momentum of the particle $a$, respectively, and $f_a$ is the distribution function of the particle $a$, which is assumed to be the Maxwell-Boltzmann distribution function. 
The collision term is calculated by
\begin{align}
   C[f_a(\bm{p}_a)] = \sum_{X,Y} C_{a + X \leftrightarrow Y}[f_a(\bm{p}_a)]~,
\end{align}
with
\begin{align}
    &C_{a + X \leftrightarrow Y}[f_a(\bm{p}_a)] \nonumber \\ 
    &=
    - \frac{1}{2 E_a} \int \prod_i \frac{\dd^3 \bm{p}_{X_i}^{}}{(2\pi)^3 2 E_{X_i}} \prod_j \frac{\dd^3 \bm{p}_{Y_j}^{}}{(2\pi)^3 2 E_{Y_j}} (2\pi)^4  \delta^{(4)}\left( p_a + \sum_i p_{X_i}^{} - \sum_j p_{Y_j}^{} \right) \nonumber \\
    & \hspace{22mm} \times \Bigg[ \sum_{\rm spins} \left| \mathcal{M}_{a + X \rightarrow Y} \right|^2 f_a(\bm{p}_a) \prod_i f_{X_i}^{}(\bm{p}_{X_i}^{}) \prod_j (1 \pm f_{Y_j}^{}(\bm{p}_{Y_j}^{}))  \nonumber \\
    & \hspace{35mm} - \sum_{\rm spins} \left| \mathcal{M}_{Y \rightarrow a + X} \right|^2 (1 \pm f_a(\bm{p}_a)) \prod_j f_{Y_j}^{}(\bm{p}_{Y_j}^{}) \prod_i (1 \pm f_{X_i}^{}(\bm{p}_{X_i}^{})) \Bigg] \nonumber \\
    &= 
    - \frac{1}{2 E_a} \int \prod_i \frac{\dd^3 \bm{p}_{X_i}^{}}{(2\pi)^3 2 E_{X_i}} \prod_j \frac{\dd^3 \bm{p}_{Y_j}^{}}{(2\pi)^3 2 E_{Y_j}} (2\pi)^4  \delta^{(4)}\left( p_a + \sum_i p_{X_i}^{} - \sum_j p_{Y_j}^{} \right) \nonumber \\
    & \hspace{22mm} \times \Bigg[ \sum_{\rm spins} \left| \mathcal{M}_{a + X \rightarrow Y} \right|^2 f_a^{\rm MB}(\bm{p}_a) \prod_i f_{X_i}^{\rm MB}(\bm{p}_{X_i}^{}) - \sum_{\rm spins} \left| \mathcal{M}_{Y \rightarrow a + X} \right|^2 \prod_j f_{Y_j}^{\rm MB}(\bm{p}_{Y_j}^{}) \Bigg]~,
\end{align}
where $f^{\rm MB}(\bm{p})$ stands for the Maxwell-Boltzmann distribution function, and the signs in the parentheses are plus for boson and minus for fermion.
The last equality is obtained by the second assumption in Sec.~\ref{sec:evolution}.

The energy transfer rates of the neutrino are calculated by
\begin{align}
\label{eq:Etransfer-nu}
   \frac{\delta \rho_\nu}{\delta t}
   &= \frac{\delta \rho_\nu}{\delta t}\Big|_{\nu \bar{\nu} \leftrightarrow e^+ e^-} 
   + \frac{\delta \rho_\nu}{\delta t}\Big|_{\nu e^\pm \leftrightarrow \nu e^\pm}  
   + \frac{\delta \rho_\nu}{\delta t}\Big|_{\phi \leftrightarrow \nu \nu}
   + \frac{\delta \rho_\nu}{\delta t}\Big|_{Z' \phi \leftrightarrow \nu \nu}~.
\end{align}
The first three terms on the right-hand side of Eq.~\eqref{eq:Etransfer-nu} are obtained by~\cite{Araki:2021xdk}
\begin{align}
   \frac{\delta \rho_\nu}{\delta t}\Big|_{\nu \bar{\nu} \leftrightarrow e^+ e^-} 
   + \frac{\delta \rho_\nu}{\delta t}\Big|_{\nu e^\pm \leftrightarrow \nu e^\pm} &=
   \frac{4 G_F^2}{\pi^5} \left\{ (g_{e L}^2 + g_{e R}^2) + 2 (g_{\mu L}^2 + g_{\mu R}^2) \right\} F(T_\gamma, T_\nu) \nonumber \\
   &\hspace{20mm} + \frac{2 (g_{Z'}^{} \epsilon e)^2}{\pi^5 m_{Z'}^4} F(T_\gamma, T_\nu)~, \\
   \frac{\delta \rho_\nu}{\delta t}\Big|_{\phi \leftrightarrow \nu \nu} &=
   \frac{m_\phi^3}{2 \pi^2} \left[ T_\nu K_2 \left( \frac{m_\phi}{T_\nu} \right) - T_\phi K_2 \left( \frac{m_\phi}{T_\phi} \right) \right] \Gamma_\phi~,
\end{align}
where $F(T_1, T_2) = 32(T_1^9 - T_2^9) + 56 T_1^4 T_2^4 (T_1 - T_2)$, $K_2(x)$ is the modified Bessel function of the second kind, and
\begin{align}
   g_{L,\alpha}^{} = \left\{ 
   \begin{array}{ll} \frac{1}{2} + \sin^2 \theta_W & (\alpha = e) \\ - \frac{1}{2} + \sin^2 \theta_W & (\alpha = \mu, \tau) \end{array} \right., \qquad
   g_{R,\alpha}^{} = \sin^2 \theta_W~,
\end{align}
with $\theta_W$ being the Weinberg mixing angle.
The energy transfer rates of the U(1)$_{L_\mu-L_\tau}$ gauge boson and Majoron are calculated by
\begin{align}
\label{eq:Etransfer-Zp}
   \frac{\delta \rho_{Z'}}{\delta t} &=
   \frac{\delta \rho_{Z'}}{\delta t}\Big|_{Z' \leftrightarrow e^- e^+} 
   + \frac{\delta \rho_{Z'}}{\delta t}\Big|_{Z' \leftrightarrow \nu \bar{\nu}} 
   + \frac{\delta \rho_{Z'}}{\delta t}\Big|_{Z' \nu \leftrightarrow \phi \bar{\nu}}
   + \frac{\delta \rho_\phi}{\delta t}\Big|_{Z' \phi \leftrightarrow \nu \nu}~, \\
\label{eq:Etransfer-phi}
   \frac{\delta \rho_\phi}{\delta t} &=
   \frac{\delta \rho_\phi}{\delta t}\Big|_{\phi \leftrightarrow \nu \nu}
   + \frac{\delta \rho_\phi}{\delta t}\Big|_{\phi \leftrightarrow \bar{\nu} \bar{\nu}}
   + \frac{\delta \rho_\phi}{\delta t}\Big|_{Z' \nu \leftrightarrow \phi \bar{\nu}}
   + \frac{\delta \rho_\phi}{\delta t}\Big|_{Z' \phi \leftrightarrow \nu \nu}~.
\end{align}
The first two terms in Eqs.~\eqref{eq:Etransfer-Zp} and \eqref{eq:Etransfer-phi} are obtained by~\cite{Araki:2021xdk}
\begin{align}
   \frac{\delta \rho_{Z'}}{\delta t}\Big|_{Z' \leftrightarrow e^- e^+} &=
   \frac{3 m_{Z'}^2}{2 \pi^2} \left[ T_e K_2 \left( \frac{m_{Z'}^{}}{T_e} \right) - T_\nu e^{\frac{2 \mu_{Z'}^{}}{T_\nu}} K_2 \left( \frac{m_{Z'}^{}}{T_\nu} \right) \right] \Gamma(Z' \to e^+ e^-) \nonumber \\
   &\approx \frac{3 m_{Z'}^2}{2 \pi^2} \left[ T_\gamma K_2 \left( \frac{m_{Z'}^{}}{T_\gamma} \right) - T_\nu K_2 \left( \frac{m_{Z'}^{}}{T_\nu} \right) \right] \Gamma(Z' \to e^+ e^-)~, \\
   \frac{\delta \rho_{Z'}}{\delta t}\Big|_{Z' \leftrightarrow \nu \bar{\nu}} &=
   \frac{3 m_{Z'}^2}{\pi^2} \left[ T_\nu e^{\frac{2 \mu_\nu}{T_\nu}} K_2 \left( \frac{m_{Z'}^{}}{T_\nu} \right) - T_\nu e^{\frac{2 \mu_{Z'}^{}}{T_{Z'}}} K_2 \left( \frac{m_{Z'}^{}}{T_\nu} \right) \right] \Gamma(Z' \to \nu_\alpha \bar{\nu}_\alpha) \nonumber \\
   &\approx 0~, \\
   \frac{\delta \rho_\phi}{\delta t}\Big|_{\phi \leftrightarrow \nu \nu} +
   \frac{\delta \rho_\phi}{\delta t}\Big|_{\phi \leftrightarrow \bar{\nu} \bar{\nu}} &=
   \frac{m_\phi^3}{2 \pi^2} \left[ T_\phi e^{\frac{\mu_\phi}{T_\phi}} K_2 \left( \frac{m_\phi}{T_\phi} \right) - T_\nu e^{\frac{\mu_\nu}{T_\nu}} K_2 \left( \frac{m_\phi}{T_\nu} \right) \right] \Gamma_\phi \nonumber \\
   &\approx \frac{m_\phi^3}{2 \pi^2} \left[ T_\phi K_2 \left( \frac{m_\phi}{T_\phi} \right) - T_\nu K_2 \left( \frac{m_\phi}{T_\nu} \right) \right] \Gamma_\phi~.
\end{align}
 
We focus on the simultaneous presence of both the U(1)$_{L_\mu-L_\tau}$ gauge boson and Majoron.
In this situation, the energy transfer derived from the $Z'\,$--$\,\phi$ interactions appears.
In contrast to the other processes, the energy transfer rates of the $Z'\,$--$\,\phi$ interactions cannot be written analytically and have to be obtained numerically by evaluating Eq.~\eqref{eq:Etransfer}.
In Subsection~\ref{subsec:matrix-element}, the invariant matrix elements of the $Z'\,$--$\,\phi$ interacting processes are shown, and a more detail discussion on the integration in the collision term is shown in Appendix~\ref{append:integ}.
In the calculation of the collision terms derived from the $Z'\,$--$\,\phi$ interaction, there is a divergence.
In Subsection~\ref{subsec:divergence}, we discuss this divergence and give the ansatz to remove this.

\subsection{Invariant matrix element}
\label{subsec:matrix-element}

In the U(1)$_{L_\mu-L_\tau} \times$ U(1)$_L$ model, there are two kinds of $Z'\,$--$\,\phi$ interactions~: Compton-like process ($Z' \nu_\alpha \leftrightarrow \phi \bar{\nu}_\beta, Z' \bar{\nu}_\alpha \leftrightarrow \phi \nu_\beta$) and pair-annihilation/creation process ($Z' \phi \leftrightarrow \nu_\alpha \nu_\beta, Z' \phi \leftrightarrow \bar{\nu}_\alpha \bar{\nu}_\beta$).
In this subsection, we show the invariant matrix elements of these processes.

\noindent {{\bf Compton-like process~:} {\boldmath $Z'\nu_\alpha \leftrightarrow \phi\bar{\nu}_\beta \quad (Z'\bar{\nu}_\alpha \leftrightarrow \phi\nu_\beta)$}}

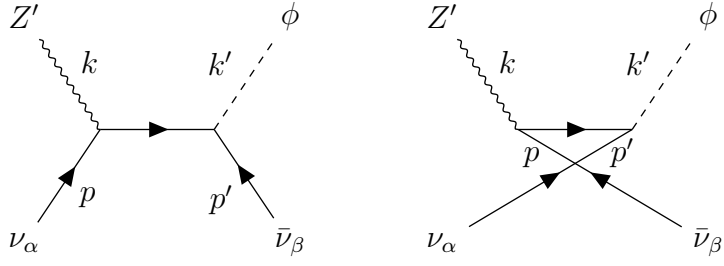
\begin{figure}[tb]
\begin{center}
\begin{tikzpicture}
\begin{feynhand}
	\vertex [particle] (fig1_i1) at (-4.5, 1.5) {$Z'$};
	\vertex [particle] (fig1_i2) at (-4.5, -1.5) {$\nu_\alpha$};
	\vertex [particle] (fig1_f1) at (-1, 1.5) {$\phi$};
	\vertex [particle] (fig1_f2) at (-1, -1.5) {$\bar{\nu}_\beta$};
	\vertex (fig1_g1) at (-3.5, 0);
	\vertex (fig1_g2) at (-2, 0);
	\propag [photon] (fig1_i1) to [edge label = \text{$k$}] (fig1_g1);
	\propag [fermion] (fig1_i2) to [edge label' = \text{$p$}] (fig1_g1);
	\propag [fermion] (fig1_g1) to (fig1_g2);
	\propag [scalar] (fig1_g2) to [edge label = \text{$k'$}] (fig1_f1);
	\propag [anti fermion] (fig1_g2) to [edge label' = \text{$p'$}] (fig1_f2);

	\vertex [particle] (fig2_i1) at (1, 1.5) {$Z'$};
	\vertex [particle] (fig2_i2) at (1, -1.5) {$\nu_\alpha$};
	\vertex [particle] (fig2_f1) at (4.5, 1.5) {$\phi$};
	\vertex [particle] (fig2_f2) at (4.5, -1.5) {$\bar{\nu}_\beta$};
	\vertex (fig2_g1) at (2, 0);
	\vertex (fig2_g2) at (3.5, 0);
	\propag [photon] (fig2_i1) to [edge label = \text{$k$}] (fig2_g1);
	\propag [fermion] (fig2_i2) to [edge label = \text{$p$}] (fig2_g2);
	\propag [fermion] (fig2_g1) to (fig2_g2);
	\propag [scalar] (fig2_g2) to [edge label = \text{$k'$}] (fig2_f1);
	\propag [anti fermion] (fig2_g1) to [edge label = \text{$p'$}] (fig2_f2);
\end{feynhand}
\end{tikzpicture}
\end{center}
\caption{$s$-channel (left) and $u$-channel (right) diagrams of the Compton-like process~: $Z'\nu_\alpha \leftrightarrow \phi\bar{\nu}_\beta$.}
\label{fig:diag_Compton}
\end{figure}

\begin{table}[tb]
\begin{center}
\begin{tabular}{c | c | c }
 & $\beta = e$ & $\beta = \mu, \tau$ \\ \hline
$\alpha = e $ & 0 & $\mathcal{M}^u$ \\ \hline
$\alpha = \mu, \tau$ & $\mathcal{M}^s$ & $\mathcal{M}^s + \mathcal{M}^u$
\end{tabular}
\caption{
Invariant matrix elements of the Compton-like reaction $Z'\nu_\alpha \leftrightarrow \phi\bar{\nu}_\beta$.
$\mathcal{M}^{s (u)}$ is derived from the $s (u)$-channel diagram shown in Fig.~\ref{fig:diag_Compton}
}
\label{tab:amp_Compton}
\end{center}
\end{table} 

As shown in Fig.~\ref{fig:diag_Compton}, two diagrams contribute to the Compton-like $Z'\,$--$\,\phi$ reaction as $Z'\nu_\alpha \leftrightarrow \phi\bar{\nu}_\beta$.
For $(\alpha, \beta) = (e, e)$, the amplitude of the Compton-like process is negligibly small because the U(1)$_{L_\mu-L_\tau}$ gauge boson does not couple to the electron neutrino at tree level.
When either the initial- or final-state neutrino is electron neutrino, and the other is mu or tau neutrino, the $s$-channel or $u$-channel diagram contributes to this process, respectively.
For $\alpha, \beta = \mu, \tau$, both the $s$- and $u$-channel diagrams contribute to the amplitude.
The relation between the neutrino flavors and contribution is summarized in Table~\ref{tab:amp_Compton}. 

The amplitude for the Compton-like process is given by
\begin{align}
\label{eq:amp_Compton}
   \sum_{\rm spins} & |\mathcal{M}_{Z'\nu_\alpha \leftrightarrow \phi\bar{\nu}_\beta}|^2 \nonumber \\
   &= \frac{16 g_\alpha^2 |h_{\alpha\beta}|^2}{(k + p)^4} \left[4 (k \cdot p) (k \cdot p') + 4 (k \cdot p) (p \cdot p') - m_{Z'}^2 (p \cdot p') + \frac{4}{m_{Z'}^2} (k \cdot p)^2 (p \cdot p') \right] \nonumber \\
   & \quad + \frac{16 g_\beta^2 |h_{\alpha\beta}|^2}{(k - p')^4} \left[4 (k \cdot p) (k \cdot p') - 4 (k \cdot p') (p \cdot p') - m_{Z'}^2 (p \cdot p') + \dfrac{4}{m_{Z'}^2} (k \cdot p')^2 (p \cdot p') \right] \nonumber \\
   & \quad + \frac{16 g_\alpha g_\beta |h_{\alpha\beta}|^2}{(k + p)^2(k - p')^2} \Big[4 (k \cdot p) (k \cdot p') - 2 (k \cdot p) (p \cdot p') + 2 (k \cdot p') (p \cdot p') + 4 (p \cdot p')^2 \nonumber \\
   & \hspace{120pt} + m_{Z'}^2 (p \cdot p') -\frac{4}{m_{Z'}^2}(k \cdot p) (k \cdot p') (p \cdot p') \Big]~,  
\end{align}
where $p$, $p'$, $k$, and $k'$ stand for the momenta of the initial-state neutrino, final-state neutrino, U(1)$_{L_\mu-L_\tau}$ gauge boson, and Majoron, respectively, and
\begin{equation}
g_\alpha = \left\{
\begin{array}{c}
0 \quad\quad (\alpha = e) \\
g_{Z'}^{} \quad (\alpha = \mu) \\
-g_{Z'}^{} \quad (\alpha = \tau) 
\end{array} \right.~.
\end{equation}

\noindent {{\bf Pair-annihilation/creation process~:}{\boldmath $Z'\phi \leftrightarrow \nu_\alpha \nu_\beta \quad (Z'\phi \leftrightarrow \bar{\nu}_\alpha \bar{\nu}_\beta)$}}

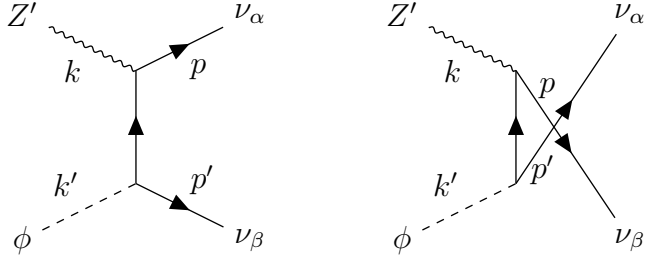
\begin{figure}[tb]
\begin{center}
\begin{tikzpicture}
\begin{feynhand}
	\vertex [particle] (fig1_i1) at (-4, 1.5) {$Z'$};
	\vertex [particle] (fig1_i2) at (-4, -1.5) {$\phi$};
	\vertex [particle] (fig1_f1) at (-1, 1.5) {$\nu_\alpha$};
	\vertex [particle] (fig1_f2) at (-1, -1.5) {$\nu_\beta$};
	\vertex (fig1_g1) at (-2.5, 0.75);
	\vertex (fig1_g2) at (-2.5, -0.75);
	\propag [photon] (fig1_i1) to [edge label' = \text{$k$}] (fig1_g1);
	\propag [scalar] (fig1_i2) to [edge label = \text{$k'$}] (fig1_g2);
	\propag [anti fermion] (fig1_g1) to (fig1_g2);
	\propag [fermion] (fig1_g1) to [edge label' = \text{$p$}] (fig1_f1);
	\propag [fermion] (fig1_g2) to [edge label = \text{$p'$}] (fig1_f2);

	\vertex [particle] (fig2_i1) at (1, 1.5) {$Z'$};
	\vertex [particle] (fig2_i2) at (1, -1.5) {$\phi$};
	\vertex [particle] (fig2_f1) at (4., 1.5) {$\nu_\alpha$};
	\vertex [particle] (fig2_f2) at (4, -1.5) {$\nu_\beta$};
	\vertex (fig2_g1) at (2.5, 0.75);
	\vertex (fig2_g2) at (2.5, -0.75);
	\propag [photon] (fig2_i1) to [edge label' = \text{$k$}] (fig2_g1);
	\propag [scalar] (fig2_i2) to [edge label = \text{$k'$}] (fig2_g2);
	\propag [anti fermion] (fig2_g1) to (fig2_g2);
	\propag [fermion] (fig2_g1) to [edge label' = \text{$p'$}, with arrow = 0.75] (fig2_f2);
	\propag [fermion] (fig2_g2) to [edge label = \text{$p$}, with arrow = 0.75] (fig2_f1);
\end{feynhand}
\end{tikzpicture}
\end{center}
\caption{$t$-channel (left) and $u$-channel (right) diagrams of the pair-annihilation/creation process~: $Z'\phi \leftrightarrow \nu_\alpha\nu_\beta$.}
\label{fig:diag_pair}
\end{figure}

\begin{table}[tb]
\begin{center}
\begin{tabular}{c | c | c }
   & $\beta = e$ & $\beta = \mu, \tau$ \\ \hline
   $\alpha = e $ & 0 & $\mathcal{M}^u$ \\ \hline
   $\alpha = \mu, \tau$ & $\mathcal{M}^t$ & $\mathcal{M}^t + \mathcal{M}^u$
\end{tabular}
\caption{
Invariant matrix elements of the pair-annihilation/creation process $Z'\phi \leftrightarrow \nu_\alpha \nu_\beta$.
$\mathcal{M}^{t (u)}$ is derived from the $s (u)$-channel diagram shown in Fig.~\ref{fig:diag_pair}.
}
\label{tab:amp_pair}
\end{center}
\end{table} 

The pair-annihilation/creation process of the U(1)$_{L_\mu-L_\tau}$ gauge boson and Majoron, $Z'\phi \leftrightarrow \nu_\alpha\nu_\beta$, is derived from the two diagrams shown in Fig.~\ref{fig:diag_pair}.
In the same way as the Compton-like process, the contribution of $t$- and $u$-channel diagrams to the amplitude depends on the neutrino flavors and is summarized in Table~\ref{tab:amp_pair}.

The amplitude for the pair-annihilation/creation process is given by
\begin{align}
   \sum_{\rm spins} & |\mathcal{M}_{Z'\phi \leftrightarrow \nu_\alpha\nu_\beta}|^2 \nonumber \\
   &= \frac{16 g_\alpha^2 |h_{\alpha\beta}|^2}{(k - p)^4} \left[4 (k \cdot p) (k \cdot p') - 4 (k \cdot p) (p \cdot p') - m_{Z'}^2 (p \cdot p') + \frac{4}{m_{Z'}^2} (k \cdot p)^2 (p \cdot p') \right] \nonumber \\
   & \quad + \frac{16 g_\beta^2 |h_{\alpha\beta}|^2}{(k - p')^4} \left[4 (k \cdot p) (k \cdot p') - 4 (k \cdot p') (p \cdot p') - m_{Z'}^2 (p \cdot p') + \frac{4}{m_{Z'}^2} (k \cdot p')^2 (p \cdot p') \right] \nonumber \\
   & \quad + \frac{16 g_\alpha g_\beta |h_{\alpha\beta}|^2}{(k - p)^2(k - p')^2} \left[4 (k \cdot p) (k \cdot p') - 2 (k \cdot p) (p \cdot p') - 2 (k \cdot p') (p \cdot p') + 4 (p \cdot p')^2 \right. \nonumber \\
   & \hspace{120pt} \left. - m_{Z'}^2 (p \cdot p') -\dfrac{4}{m_{Z'}^2}(k \cdot p) (k \cdot p') (p \cdot p') \right]~,
\end{align}
where $p\,(p')$, $k$, and $k'$ stand for the momenta of the neutrino with the flavor $\alpha\,(\beta)$, U(1)$_{L_\mu-L_\tau}$ gauge boson, and Majoron, respectively.

\subsection{Divergence of the \texorpdfstring{$u$}{u}-channel diagram of the Compton-like process}
\label{subsec:divergence}

In the calculation of the energy transfer rate, the amplitude is integrated over the momentum of the interested particle. 
However, for the $u$-channel diagram of the Compton-like process, the denominator of the propagator of the intermediate state becomes zero at a certain momentum, and then, the integration of the amplitude diverges.
This divergence is interpreted as that the intermediate state becomes a real particle at this momentum, i.e., an on-shell neutrino is generated by a decay of the U(1)$_{L_\mu-L_\tau}$ gauge boson $Z'\to \nu \bar{\nu}$, and it induces an inverse decay of the Majoron $\nu\nu \to \phi$.
This phenomenon is caused by the presence of unstable particles in the initial states.

This divergent contribution must be removed because it has already been included in the energy transfer rates of the $Z'$ decay and $\phi$ inverse decay.
A similar problem appears in the context of leptogenesis~\cite{Fukugita:1986hr}.
In the calculation of the lepton asymmetry produced in the leptogenesis scenario, the lepton number violating two-to-two scattering appears, and the contribution from on-shell heavy Majorana neutrinos has to be subtracted.
For this case, the heavy Majorana neutrinos are unstable particles, and their propagators are regulated by their decay widths.
The on-shell contribution is subtracted based on these decay widths (for instance, see Ref.~\cite{Buchmuller:2004nz}).
On the other hand, in our model, the propagating active neutrinos are stable, and there is no indicator of the width to remove the double counting in the original integrand of the collision term.

As regularizations of this divergence, the following methods have been proposed.
In Ref.~\cite{Grzadkowski:2021kgi}, all particles that propagate while interacting with a gas can be considered ``quasiparticles'' because they have finite mean free paths.
There is an imaginary contribution to the self-energy of the particles in the medium, and this imaginary part plays the role of the decay width of the quasiparticle and can regularize the divergence.
However, since Ref.~\cite{Grzadkowski:2021kgi} has concluded that this method is valid only in a high-energy regime, it cannot be used to track the time evolution of the universe.
Another proposal given in Ref.\cite{Ginzburg:1995bc} is to regulate the divergence by setting the mass of the unstable particle in the initial state to $m \to m - i\Gamma$ with $\Gamma$ being its decay width.
However, there is a fatal problem that the energy conservation at a vertex of the final state is no longer valid. 

Here, we propose the following prescription to avoid these problems. 
For the amplitude of the $u$-channel diagram
\begin{equation*}
   |\mathcal{M}_{Z'\nu_\alpha \leftrightarrow \phi\bar{\nu}_\beta}^u|^2 \propto 
   \frac{1}{(m_{Z'}^2 - 2 k \cdot p')^2}~, 
\end{equation*}
we consider the following integral by transforming an integration variable as $2 k \cdot p' = X$~:
\begin{equation}
   I = 
   \int^\infty_0 \dd X \frac{f(X)}{(m_{Z'}^2 - X)^2 + \varepsilon^2}~. 
\end{equation} 
The above transformation allows the divergence to depend only on $X$.
Since the integrand has a divergence of $\mathcal{O}(1/\varepsilon^2)$ at $X = m_{Z'}^2$, and the width of the divergence is about $\varepsilon$, 
we can expand the result of the integration over $X$ by $\varepsilon$ as follows~:
\begin{equation}
   I = 
   \frac{\alpha_{-1}}{\varepsilon} + \alpha_0 + \varepsilon\alpha_1 + \cdots~.
\label{epsilon expand}
\end{equation}
Here, the first term diverges as $\varepsilon \to 0$, and hence, it comes from the on-shell mediator contribution.
On the other hand, the second term comes from the off-shell one. The higher-order terms do not exist originally because they disappear in the limit $\varepsilon \rightarrow 0$. Therefore, the off-shell contribution is obtained from $\alpha_0$. 
In our calculation, we extract the off-shell contribution $\alpha_0$ to evaluate the energy transfer rate of the $Z'\,$--$\,\phi$ interactions.

\section{Results}
\label{sec:result}

In this section, we display the deviation of the effective number of neutrino species from the SM prediction for the cases with and without the $Z'\,$--$\,\phi$ interactions and show the parameter regions of the Majoron with the U(1)$_{L_\mu-L_\tau }$ gauge boson that can alleviate the Hubble tension.
In the calculation of the collision terms, we use VEGAS~\cite{Lepage:1977sw,peter_lepage_2023_8175999} for performing multiple integrals.

\subsection{Initial conditions of Majoron and its parameters}
\label{subsec:parameter}

In this paper, we focus on the Majoron contribution to $N_{\rm eff}$ in the presence of the U(1)$_{L_\mu-L_\tau}$ gauge boson.
Because of the tiny Majoron coupling to the neutrinos, the Majorons never reach thermal equilibrium and are slowly produced in the early universe.
The production rate of the Majoron has a peak at $T_\nu/m_\phi \sim 1/3$ and becomes negligibly small at $T_\nu/m_\phi \lesssim 0.05$~\cite{EscuderoAbenza:2020cmq}.
For the enough large contribution to $N_{\rm eff}$ from the Majorons,  the Majorons have to decay before the neutrino decoupling, which occurs at $T_\gamma = 0.5$\,MeV~\cite{Escudero:2019gzq}.
Therefore, the Majoron mass needs to satisfy
\begin{equation*}
   T_\nu \sim 0.05 m_\phi \lesssim 0.5\,{\rm MeV} \quad \Rightarrow \quad m_\phi \lesssim 10\,{\rm MeV}~,
\end{equation*}
in order to contribute to $N_{\rm eff}$.

For $T_\nu/m_\phi \lesssim 5$, the ratio of the Majoron production rate to the Hubble parameter, $\langle \Gamma_{\nu\nu \to \phi} \rangle/H$, is smaller than one, and the Majoron density would be negligible even for the case where the peak value of the ratio,
\begin{equation}
   \Gamma_{\rm eff} \equiv \left. \frac{\langle \Gamma_{\nu\nu \to \phi} \rangle}{H} \right|_{T_\nu = m_\phi/3} \simeq \left( \frac{\lambda}{4.0 \times 10^{-12}} \right)^2 \left( \frac{\rm keV}{m_\phi} \right)~,
\end{equation}
is much larger than one.
Moreover, when the initial condition of the Majoron temperature is set to be less than 1\,MeV, the initial condition does not depend on the value of $N_{\rm eff}$.
Therefore, we set the initial condition of the temperatures of the photon, neutrino, and Majoron to solve the Boltzmann equations as follows~:
\begin{equation}
   T_{0\gamma} = T_{0\nu} = 
   50\,{\rm MeV}, \quad T_{0\phi} = 1\,{\rm MeV}~.
\end{equation}

\subsection{Comparison between the cases with and without scattering}
\label{subsec:comparison}

Here, we compare the cases with and without the $Z'\,$--$\,\phi$ interactions.
The parameters of the Majoron in the calculation are chosen as 
\begin{align}
m_\phi &= 0.05, \; 0.1, \; 0.5, \; 1.0, \; 5.0, \; 10.0 \; [{\rm MeV}]~,  \\
\Gamma_{\rm eff} &= 0.01, \; 0.1, \; 1.0, \; 10, \; 100~,
\end{align}
and the mass of U(1)$_{L_\mu-L_\tau}$ gauge boson and gauge coupling are taken to be $m_{Z'}^{} = 13$\,MeV and $g_{Z'}^{} = 5.0 \times 10^{-4}$, which can explain the muon $g-2$ anomaly at $2\sigma$ level.

\begin{table}[tb]
\centering
\begin{tabular}{c c c c c c c}
\multicolumn{3}{c}{$m_\phi=0.05 \; {\rm MeV}$} & \hspace{1.0cm} & \multicolumn{3}{c}{$m_\phi=0.1 \; {\rm MeV}$} \\ \cline{1-3} \cline{5-7}
$\Gamma_{\rm eff}$ & without scat. & with scat. & \hspace{1.0cm} & $\Gamma_{\rm eff}$ & without scat. & with scat. \\ \cline{1-3} \cline{5-7}
0.01 & 3.45399  & 3.45364 & \hspace{1.0cm} & 0.01 & 3.45419 & 3.45381 \\
0.1 & 3.50778 & 3.50757 & \hspace{1.0cm} & 0.1 & 3.50838 & 3.50838 \\
1.0 & 3.55920 & 3.55921 & \hspace{1.0cm} & 1.0 & 3.56193 & 3.56193 \\
10 & 3.59129 & 3.59128 & \hspace{1.0cm} & 10 & 3.61269 & 3.61275 \\
100 & 3.62928 & 3.62929 & \hspace{1.0cm} & 100 & 3.79043 & 3.79066 \\ \cline{1-3} \cline{5-7}
\\
\multicolumn{3}{c}{$m_\phi=0.5 \; {\rm MeV}$} & \hspace{1.0cm} & \multicolumn{3}{c}{$m_\phi=1.0 \; {\rm MeV}$} \\ \cline{1-3} \cline{5-7}
$\Gamma_{\rm eff}$ & without scat. & with scat. & \hspace{1.0cm} & $\Gamma_{\rm eff}$ & without scat. & with scat. \\ \cline{1-3} \cline{5-7}
0.01 & 3.45843 & 3.45801 & \hspace{1.0cm} & 0.01 & 3.46957 & 3.46911 \\
0.1 & 3.53508 & 3.53453 & \hspace{1.0cm} & 0.1 & 3.62285 & 3.62225 \\
1.0 & 3.75503 & 3.75454 & \hspace{1.0cm} & 1.0 & 3.90820 & 3.90763 \\
10 & 4.12994 & 4.12947 & \hspace{1.0cm} & 10 & 3.87105 & 3.87057 \\
100 & 4.12709 & 4.12656 & \hspace{1.0cm} & 100 & 3.86469 & 3.86413 \\ \cline{1-3} \cline{5-7}
\\
\multicolumn{3}{c}{$m_\phi=5.0 \; {\rm MeV}$} & \hspace{1.0cm} & \multicolumn{3}{c}{$m_\phi=10.0 \; {\rm MeV}$} \\ \cline{1-3} \cline{5-7}
$\Gamma_{\rm eff}$ & without scat. & with scat. & \hspace{1.0cm} & $\Gamma_{\rm eff}$ & without scat. & with scat. \\ \cline{1-3} \cline{5-7}
0.01 & 3.43964 & 3.43949 & \hspace{1.0cm} & 0.01 & 3.39982 & 3.39927 \\
0.1 & 3.40058 & 3.40021 & \hspace{1.0cm} & 0.1 & 3.39943 & 3.39904 \\
1.0 & 3.40027 & 3.39982 & \hspace{1.0cm} & 1.0 & 3.39944 & 3.39899 \\
10 & 3.40026 & 3.39980 & \hspace{1.0cm} & 10 & 3.39945 & 3.39893 \\
100 & 3.40025 & 3.39984 & \hspace{1.0cm} & 100 & 3.39944 & 3.39899 \\ \cline{1-3} \cline{5-7}
\\
\end{tabular}
\caption{Comparison of $N_{\rm eff}$ calculated with and without the $Z'\,$--$\,\phi$ interactions.}
\label{tab:Neff-comparison}
\end{table}

In Tab.~\ref{tab:Neff-comparison}, the values of $N_{\rm eff}$ for the cases with and without the $Z'\,$--$\,\phi$ interactions are summarized. 
By comparing the values of $N_{\rm eff}$, it is shown that there are only $\mathcal{O}(10^{-4})$ changes even when the $Z'\,$--$\,\phi$ interactions are taken into account, and $N_{\rm eff}$ tends to be a little smaller.
This is because the initial temperature of the Majorons $T_{0 \phi}$ is low, and the energy flows from the neutrinos and $Z'$ to the Majorons by the $Z'\,$--$\,\phi$ interactions. 
Since the contribution of the $Z'$ decay to $N_{\rm eff}$ is larger than that of the Majoron decay, a small decrease in $Z'$ results in a smaller $N_{\rm eff}$.
From this result, the difference between the cases with and without considering the $Z'\,$--$\,\phi$ interactions is not significant from the viewpoint of the alleviation of the Hubble tension.

\subsection{Parameter regions of Majoron}
\label{subsec:majoron-parameter}

In this subsection, we explore the Majoron contribution to $N_{\rm eff}$ and show the favored and excluded regions of the Majoron parameters.
By combining the results from the CMB, Cepheids, and others, $N_{\rm eff}$ is derived as $3.27 \pm 0.15$ at $68\%$ C.L.~\cite{Planck:2018vyg}, which implies that $0.3 \lesssim \Delta N_{\rm eff} \lesssim 0.5$ with $\Delta N_{\rm eff} \equiv \Delta N_{\rm eff}^{Z'} + \Delta N_{\rm eff}^{\phi} + \Delta N_{\rm eff}^{Z'\phi}$ needs to alleviate the Hubble tension.
However, large $\Delta N_{\rm eff}$ changes the expansion rate of the universe during Big Bang Nucleosynthesis (BBN) and spoils the successful generation of light elements.
In Ref.~\cite{Pitrou:2018cgg}, the constraint on $N_{\rm eff}$ from BBN is given as $N_{\rm eff} = 2.88 \pm 0.27$ (68\% C.L.), and we adopt $N_{\rm eff} < 3.5$ as the one-sided 95\% C.L. upper limit of $N_{\rm eff}$, following Ref.~\cite{Escudero:2019gvw}.

For focusing on the contribution of the Majoron to $N_{\rm eff}$, we fix the parameters of the U(1)$_{L_\mu-L_\tau}$ gauge boson which can explain the muon $g-2$ anomaly at $2\sigma$ level.
The effective number of neutrino species in the absence of the Majoron, $N_{\rm eff}^{\rm SM} + \Delta N_{\rm eff}^{Z'}$, is evaluated as
\begin{equation}
\label{eq:Neff-Zponly}
    N_{\rm eff}^{\rm SM} + \Delta N_{\rm eff}^{Z'} \simeq 
    \left\{ \begin{array}{l}
       3.4 \\   3.2 \\   3.043
    \end{array} \right. \text{for } (m_{Z'}^{}, g_{Z'}^{}) = \left\{ 
    \begin{array}{l}
       (13\,{\rm MeV}, 5.0\times 10^{-4})\\   (18\,{\rm MeV}, 4.0 \times 10^{-4}) \\    (100\,{\rm MeV}, 7.0 \times 10^{-4})
    \end{array}
    \right.~.
\end{equation}
For the case of $(m_{Z'}^{}, g_{Z'}^{}) = (100\,{\rm MeV, 7.0 \times 10^{-4}})$, $\Delta N_{\rm eff}^{Z'}$ is negligibly small.
This is because too heavy $Z'$ decays before the neutrino decoupling and does not contribute to $N_{\rm eff}$.

From the results of the previous subsection, the contribution of the $Z'\,$--$\,\phi$ interactions to $N_{\rm eff}$ is small enough, and therefore, we do not incorporate the $Z'\,$--$\,\phi$ interactions in the following calculations.
The initial conditions for temperatures are the same as those in subsection~\ref{subsec:parameter}, and we calculate $N_{\rm eff}$ for the three cases of the $Z'$ parameters in Eq.~(\ref{eq:Neff-Zponly}).
\begin{figure}[!t]
\begin{minipage}[b]{0.49\linewidth}
   \centering
   \includegraphics[keepaspectratio, scale=0.49]{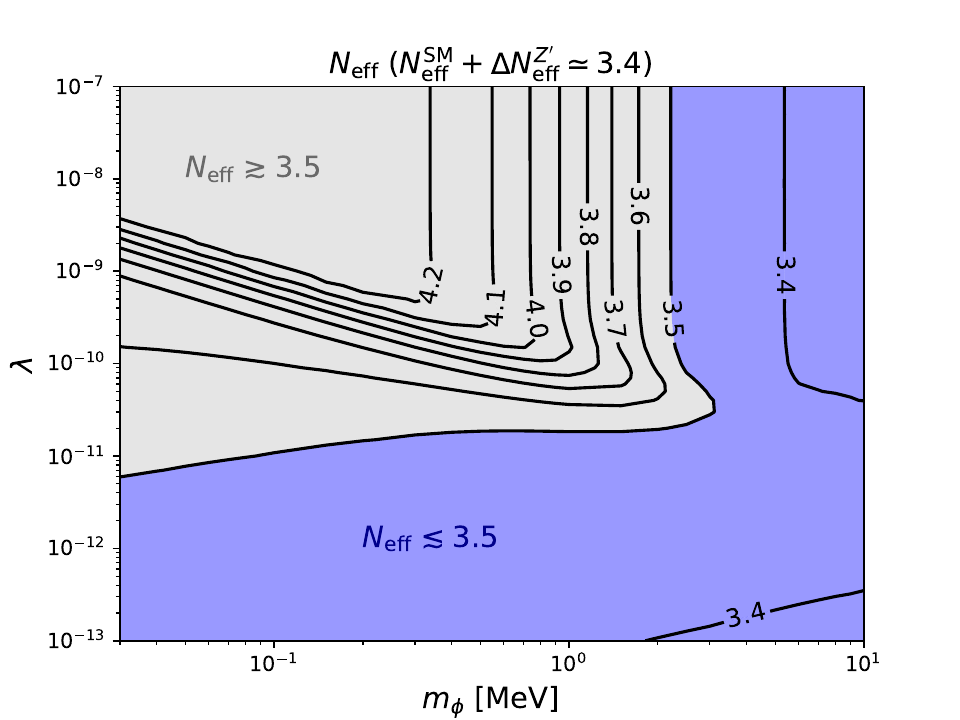}
    \subcaption{$(m_{Z'}^{}, g_{Z'}^{}) = (13\,{\rm MeV}, 5.0 \times 10^{-4})$}
\end{minipage}
\begin{minipage}[b]{0.49\linewidth}
   \centering
   \includegraphics[keepaspectratio, scale=0.49]{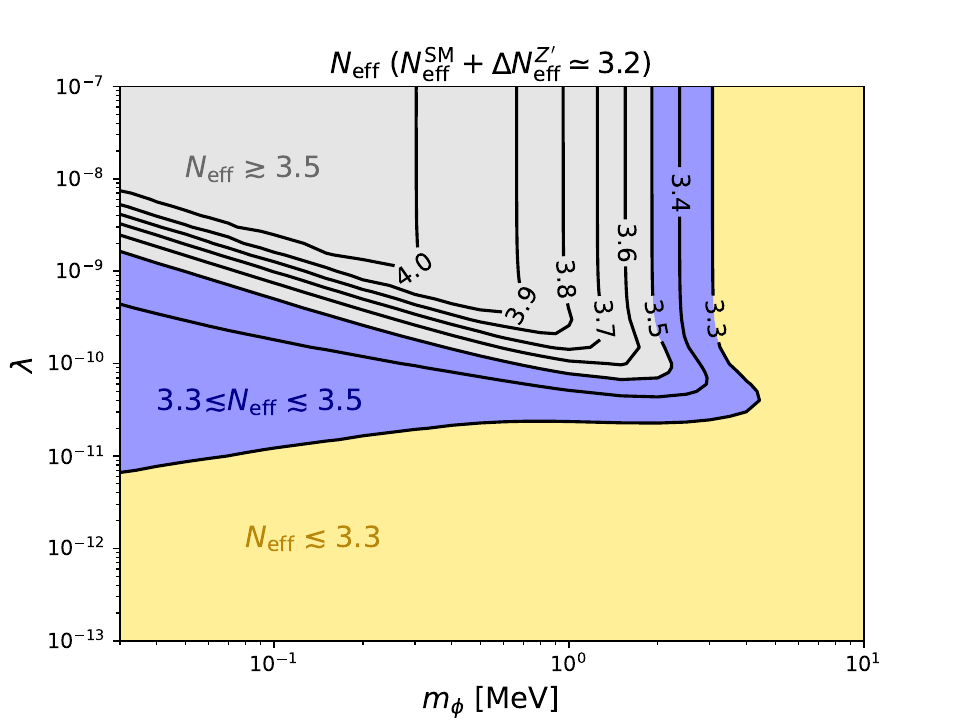}
    \subcaption{$(m_{Z'}^{}, g_{Z'}^{}) = (18\,{\rm MeV}, 4.0 \times 10^{-4})$}
\end{minipage} \\
\begin{minipage}[b]{1.0\linewidth}
   \centering
   \includegraphics[keepaspectratio, scale=0.49]{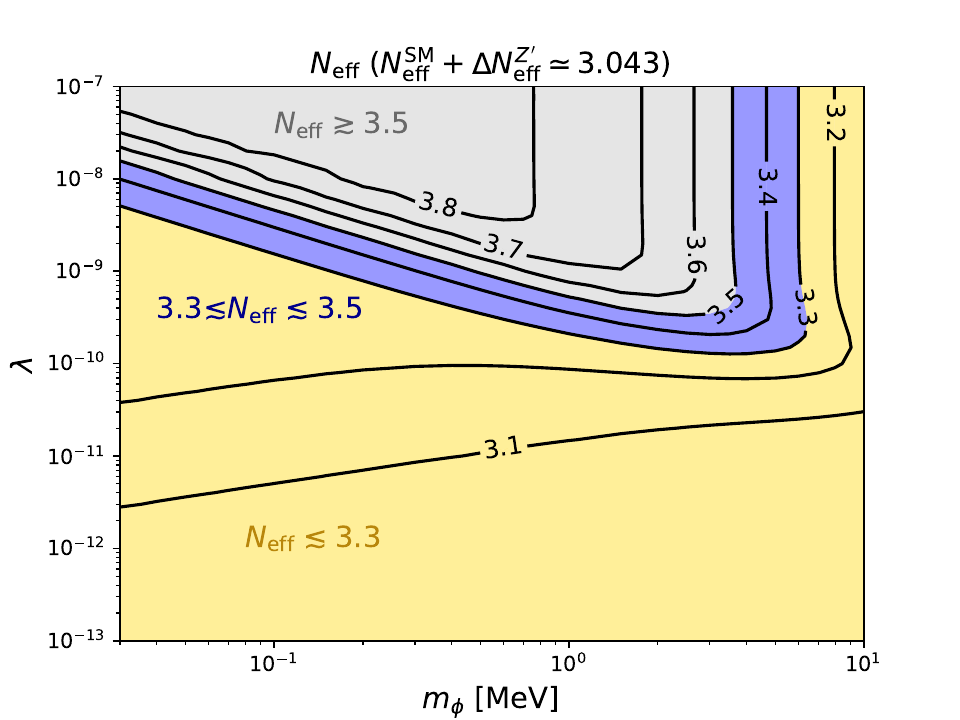}
    \subcaption{$(m_{Z'}^{}, g_{Z'}^{}) = (100\,{\rm MeV}, 7.0 \times 10^{-4})$}
\end{minipage} 
\caption{
Contours of $N_{\rm eff}$ for $(m_{Z'}^{}, g_{Z'}^{}) = (13\,{\rm MeV}, 5.0 \times 10^{-4}), (18\,{\rm MeV}, 4.0 \times 10^{-4})$, and $(100\,{\rm MeV}, 7.0 \times 10^{-4})$, which are corresponding to $N_{\rm eff}^{\rm SM} + N_{\rm eff}^{Z'} \simeq 3.4, 3.2$, and $3.043$, respectively.
The blue shaded regions can alleviate the Hubble tension, and the gray shaded regions are excluded by BBN~\cite{Pitrou:2018cgg}.
}
\label{fig:Neff}
\end{figure}
Figure~\ref{fig:Neff} shows the contours of $N_{\rm eff}$ in the plane of the Majoron parameter ($m_\phi, \lambda$) for $(m_{Z'}^{}, g_{Z'}^{}) = (13\,{\rm MeV, 5.0 \times 10^{-4}}), (18\,{\rm MeV, 4.0 \times 10^{-4}})$, and $(100\,{\rm MeV, 7.0 \times 10^{-4}})$.

As shown in Fig.~\ref{fig:Neff}, the Hubble tension can be alleviated roughly in the following regions~:
\begin{itemize}
\item $(m_{Z'}^{}, g_{Z'}^{}) = (13\,{\rm MeV, 5.0 \times 10^{-4}}) \quad [N_{\rm eff}^{\rm SM} + N_{\rm eff}^{Z'} \simeq 3.4]$ \\[5pt]
   \qquad $m_\phi \lesssim 3$\,MeV and $\lambda \lesssim \mathcal{O}(10^{-11})$~, \\[5pt]
   \qquad $m_\phi \gtrsim 3$\,MeV~.
\item $(m_{Z'}^{}, g_{Z'}^{}) = (18\,{\rm MeV, 4.0 \times 10^{-4}}) \quad [N_{\rm eff}^{\rm SM} + N_{\rm eff}^{Z'} \simeq 3.2]$ \\[5pt]
   \qquad $m_\phi \lesssim 2$\,MeV and $\lambda \simeq \mathcal{O}(10^{-11} - 10^{-9})$~, \\[5pt]
   \qquad $2\,{\rm MeV} \lesssim m_\phi \lesssim 3$\,MeV and $\lambda \gtrsim \mathcal{O}(10^{-11})$~.
\item $(m_{Z'}^{}, g_{Z'}^{}) = (100\,{\rm MeV, 7.0 \times 10^{-4}}) \quad [N_{\rm eff}^{\rm SM} + N_{\rm eff}^{Z'} \simeq N_{\rm eff}^{\rm SM} \simeq 3.043]$ \\[5pt]
   \qquad $m_\phi \lesssim 4$\, MeV and $\lambda \simeq \mathcal{O}(10^{-10}-10^{-8})$~, \\[5pt]
   \qquad $4\,{\rm MeV} \lesssim m_\phi \lesssim 7$\,MeV and $\lambda \gtrsim \mathcal{O}(10^{-10})$~.
\end{itemize}
It is shown that the contours are vertical and do not depend on the coupling in the larger coupling region~$\lambda \gtrsim 10^{-10}$.
This is because, for the large coupling, the Majoron is in the thermal equilibrium with the SM particles in the early universe, and the contribution of the Majoron to $N_{\rm eff}$ depends only on whether most of the Majorons decay after the neutrino decoupling or not.

As shown in Fig.~\ref{fig:Neff}, the contours move toward the larger $\lambda$ and smaller $m_\phi$ as $Z'$ becomes heavier ($\Delta N_{\rm eff}^{Z'}$ becomes smaller). 
By focusing on the same $\Delta N_{\rm eff}^\phi + \Delta N_{\rm eff}^{Z'\phi}$ values, the vertical shift can be explained as follows~:
\begin{enumerate}
\item Heavier $Z'$ decays faster, and hence, less energy is injected into the neutrino sector after the neutrino decoupling. 
\item The temperature of the neutrinos $T_\nu$ after the $Z'$ decays becomes lower.
\item The length of the period where the rate of the inverse decay $\nu\nu \to \phi$ exceeds the Hubble parameter becomes shorter.
\item A larger $\lambda$ is required to produce the same amount of Majoron in the shorter time.
\end{enumerate}
In addition, the horizontal shift can be explained as follows~:
\begin{enumerate}
\item Heavier $Z'$ makes the time where the reaction rate of $e^+e^- \rightarrow Z'$ becomes smaller than the Hubble parameter earlier.
\item The temperature of the neutrino decoupling increases\footnote{
We checked this by the numerical calculations.
}, and more Majorons decay after the neutrino decoupling.
\item As a result, $\Delta N_{\rm eff}$ becomes larger.
\end{enumerate}

We also mention that there is a bump derived from an enhancement of the Majoron contribution to $N_{\rm eff}$ around $\lambda \sim 10^{-10}$ and $m_\phi = \mathcal{O}(1)$\,MeV in Fig.~\ref{fig:Neff}.
This behavior comes from the fact that because of the expansion of the universe, the energy density of the massive dark photon is relatively enhanced by the time of the decay in comparison with that of the other radiation.\footnote{
This kind of behavior also appears in the dark photon model~\cite{Ibe:2019gpv}.
}

\begin{figure}[!t]
\begin{minipage}[b]{0.49\linewidth}
   \centering
   \includegraphics[keepaspectratio, scale=0.49]{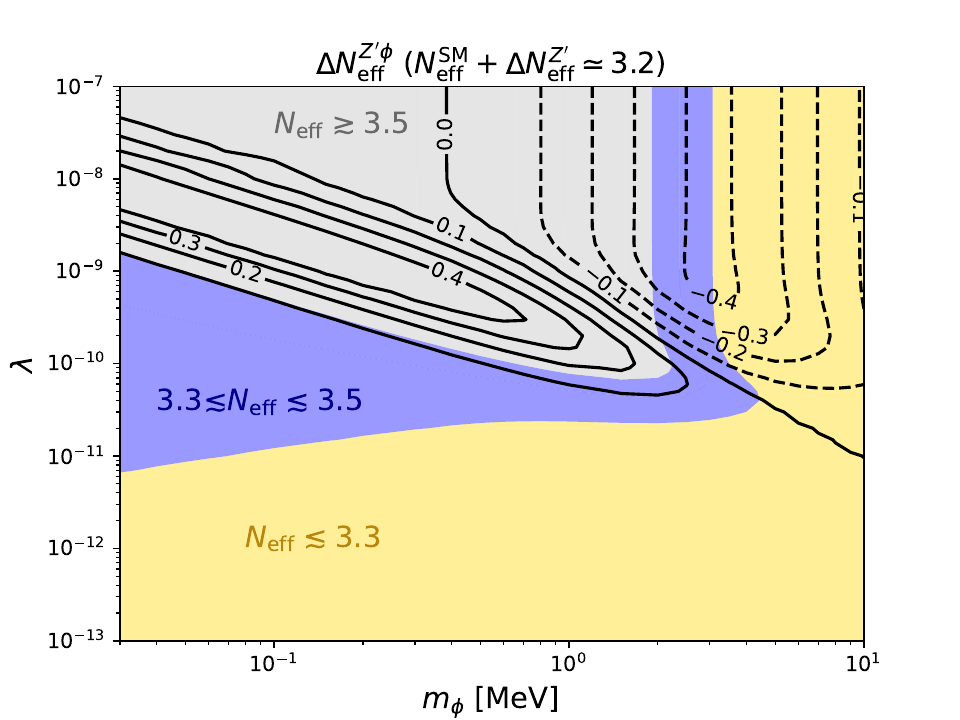}
    \subcaption{$(m_{Z'}^{}, g_{Z'}^{}) = (13\,{\rm MeV}, 5.0 \times 10^{-4})$}
\end{minipage}
\begin{minipage}[b]{0.49\linewidth}
   \centering
   \includegraphics[keepaspectratio, scale=0.49]{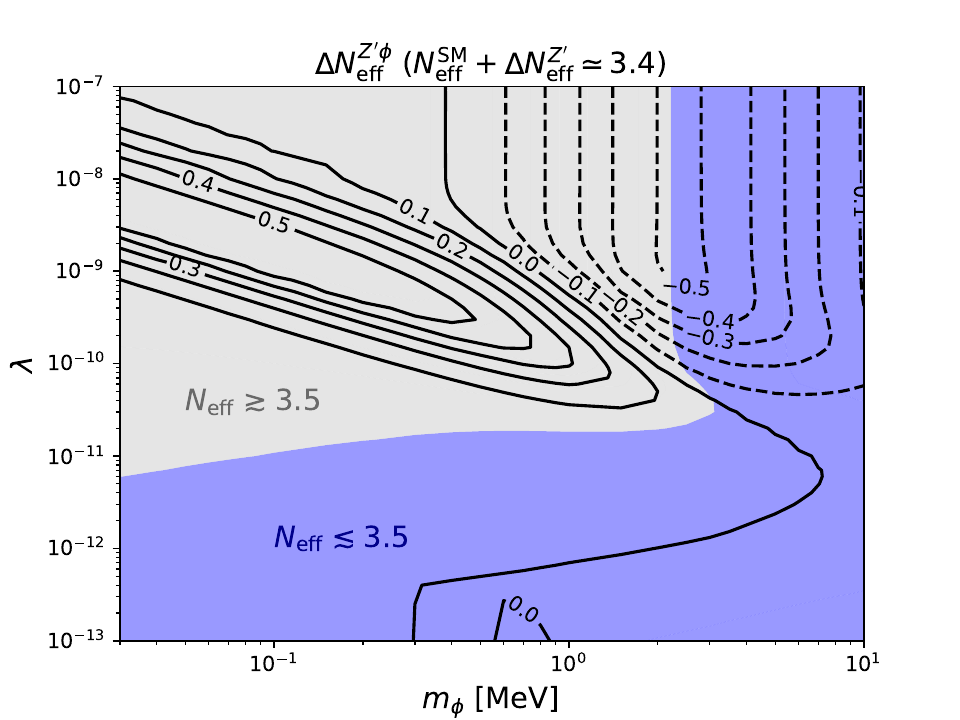}
    \subcaption{$(m_{Z'}^{}, g_{Z'}^{}) = (18\,{\rm MeV}, 4.0 \times 10^{-4})$}
\end{minipage}
\caption{
Contours of the synergy contribution between the U(1)$_{L_\mu-L_\tau}$ gauge boson and Majoron, $\Delta N_{\rm eff}^{Z'\phi}$, for $(m_{Z'}^{}, g_{Z'}^{}) = (13\,{\rm MeV}, 5.0 \times 10^{-4})$ and $(18\,{\rm MeV}, 4.0 \times 10^{-4})$, which are corresponding to $N_{\rm eff}^{\rm SM} + N_{\rm eff}^{Z'} \sim 3.2$ and $3.4$, respectively.
The blue shaded regions can alleviated the Hubble tension, and the gray shaded regions are excluded by BBN~\cite{Pitrou:2018cgg}.
}
\label{fig:Neff-Zphi}
\end{figure}

Lastly, we discuss the synergy contribution between the U(1)$_{L_\mu-L_\tau}$ gauge boson and Majoron, $\Delta N_{\rm eff}^{Z'\phi}$.
Figure~\ref{fig:Neff-Zphi} shows the Contours of the synergy contribution between the U(1)$_{L_\mu-L_\tau}$ gauge boson and Majoron, $\Delta N_{\rm eff}^{Z'\phi}$, for $(m_{Z'}^{}, g_{Z'}^{}) = (13\,{\rm MeV}, 5.0 \times 10^{-4})$ and $(18\,{\rm MeV}, 4.0 \times 10^{-4})$, which are corresponding to $N_{\rm eff}^{\rm SM} + N_{\rm eff}^{Z'} \sim 3.2$ and $3.4$, respectively.
For $(m_{Z'}^{}, g_{Z'}^{}) = (100\,{\rm MeV}, 7.0 \times 10^{-4})$, there is no synergy contribution because of too heavy mass and too early leaving of the U(1)$_{L_\mu-L_\tau}$ gauge boson. 
As shown in Fig.~\ref{fig:Neff-Zphi}, the  U(1)$_{L_\mu-L_\tau}$ gauge boson and Majoron give a negative contribution in heavy Majoron mass ($m_\phi \gtrsim 1$\,MeV) and large coupling ($\lambda \gtrsim 10^{-10}$) region.
This negative contribution comes from the fact that the more Majorons with $\mathcal{O}(1)$\,MeV mass decay before the neutrino decoupling because the light U(1)$_{L_\mu-L_\tau}$ gauge boson makes the neutrino decoupling later.
On the other hand, they give a positive contribution in light Majoron mass ($m_\phi \lesssim 1$\,MeV) and large coupling ($10^{-10} \lesssim \lambda \lesssim 10^{-7}$) region.
For the light $m_{Z'}$ case, the decays of the U(1)$_{L_\mu-L_\tau}$ gauge bosons make the neutrino temperature $T_\nu$ higher, and the length of the period where the rate of the inverse decay $\nu\nu \to \phi$ exceeds the Hubble parameter becomes longer.
Therefore, $\Delta N_{\rm eff}$ becomes larger in comparison with the case without the U(1)$_{L_\mu-L_\tau}$ gauge boson.
For the smaller and larger Majoron coupling regions, the synergy contribution decreases.
This is because the less Majorons are produced in the smaller coupling region ($\lambda \lesssim 10^{-10}$) and give the smaller contribution to $N_{\rm eff}$.
On the contrary, in the larger coupling region ($\lambda \gtrsim 10^{-7}$), the Majoron is in the thermal equilibrium with the SM particles, and the effect of the earlier neutrino decoupling by the U(1)$_{L_\mu-L_\tau}$ gauge boson can be neglected.
Therefore, in the larger coupling region, the synergy contribution $\Delta N_{\rm eff}^{Z'\phi}$ becomes smaller, and the contribution from the Majoron $\Delta N_{\rm eff}^\phi$ conversely becomes larger, as shown in Fig.~\ref{fig:Neff}.

\section{Summary}
\label{sec:summary}

In this paper, we analyze parameter regions of the Majoron that can alleviate the Hubble tension in the U(1)$_{L_\mu - L_\tau}$ model with the broken lepton number U(1)$_L$ symmetry.
As new particles, this model has a U(1)$_{L_\mu - L_\tau}$ gauge boson $Z'$ and a Majoron $\phi$, which can affect the effective number of neutrino species $N_{\rm eff}$. 
If $Z'$ and $\phi$ exist simultaneously in the early universe, $Z'\,$--$\,\phi$ interaction processes, such as $Z'\nu_\alpha \leftrightarrow \phi\bar{\nu}_\beta$, occur. 

The comparison of $N_{\rm eff}$ between the cases with and without the $Z'\,$--$\,\phi$ interaction processes shows that these processes give a negligibly small contribution of $\mathcal{O}(10^{-4})$ to $N_{\rm eff}$, and it does not need to be considered for the alleviation of the Hubble tension.
Based on these facts, we calculated $N_{\rm eff}$ for various Majoron parameters without considering the $Z'\,$--$\,\phi$ interaction processes and search parameters that could alleviate the Hubble tension. 
As a result, we found that the Majoron can alleviate the Hubble tension in some regions, which depend on the $Z'$ parameters, and the parameter region with a lighter mass $m_\phi \lesssim 2$\,MeV and a larger coupling $\lambda \gtrsim 10^{-8}$ is excluded because it predicts $N_{\rm eff} \gtrsim 3.5$.
Moreover, we also examined the synergy contribution between the U(1)$_{L_\mu-L_\tau}$ gauge boson and Majoron, $\Delta N_{\rm eff}^{Z'\phi}$, and found that the U(1)$_{L_\mu-L_\tau}$ gauge boson and Majoron give the non-trivial contribution to $N_{\rm eff}$ which does not appear until both are included in one model.
The favored and restricted regions of the Majoron parameters depend on the $Z'$ parameters because of the presence of the $Z'$ contribution and synergy one.

\section*{Acknowledgments}

The authors thank Makiko Nio for the useful discussion and advice on Monte Carlo integration using VEGAS~\cite{Lepage:1977sw,peter_lepage_2023_8175999}.
Numerical calculations are conducted on RSCC and RICC supercomputer systems at RIKEN.
The Feynman diagrams in this paper were drawn using TikZ-FeynHand~\cite{Ellis:2016jkw,Dohse:2018vqo}.
This work was supported by JSPS KAKENHI Grant Numbers JP21K20365 [KA], JP23K13097 [KA], and JP18H01210 [JS, MJSY], and MEXT KAKENHI Grant Number JP18H05543 [JS, MJSY].

\appendix

\section{Integration in collision term}
\label{append:integ}

Here, following the method of Refs.~\cite{Kreisch:2019yzn, Fradette:2018hhl, Hannestad:1995rs}, we show the collision term for the reaction process of $1+2 \leftrightarrow 3+4$,
\begin{align}
   C[f_1] &= 
   -\dfrac{1}{2E_1}\mathcal{I}~, \\
   \mathcal{I} &\equiv 
   \int \dd\Pi_2 \dd\Pi_3 \dd\Pi_4 \Lambda(\{f_i\})(2\pi)^4\delta^{(4)}(p_1 + p_2 - p_3 - p_4) \sum_\mathrm{spins} |\mathcal{M}|^2~,
\end{align}
with $\dd\Pi_i = \dd^3 {\bm p}_i / (2\pi)^3 / 2E_i$ and $\Lambda(\{f_i\}) = f_1f_2(1 \pm f_3)(1 \pm f_4) - f_3 f_4 (1 \pm f_1)(1 \pm f_2)$,
is transformed into a quadruple integral.
First, by using a formula for Lorentz-invariant integral measures,
\begin{equation}
   \frac{\dd^3 {\bm p}_i}{2E_i} =
   \dd^4 p_i \delta(p_i^2 - m_i^2) \Theta(p_i^0)~, 
\end{equation}
with $\delta(x)$ and $\Theta(x)$ being the Dirac delta function and the Heaviside step function, respectively, the $p_4$ integral is performed as
\begin{equation}
   \mathcal{I} = 
   \int \dd \Pi_2 \dd \Pi_3 \Lambda(\{f_i\}) (2\pi) \delta(p_4^2 - m_4^2) \Theta(p_4^0) \sum_\mathrm{spins} \left. |\mathcal{M}|^2 \right|_{p_4 = p_1 + p_2 - p_3}~. 
\label{eq:collision-term}
\end{equation}

Next, by taking a coordinate system such that ${\bm p}_1$ points to the $z$-axis, the momenta are expressed as
\begin{align}
   p_1 &= 
   (E_1, 0, 0, |{\bm p}_1|) \, , \\
   p_2 &= 
   (E_2, |{\bm p}_2| \sin \alpha \cos \beta~, |{\bm p}_2| \sin \alpha \sin \beta~, |{\bm p}_2| \cos \alpha)~,  \\
   p_3 &= 
   (E_3, |{\bm p}_3| \sin \theta \cos \mu~, |{\bm p}_3| \sin \theta \sin \mu~, |{\bm p}_3| \cos \theta)~,  \\
   p_4 &= 
   p_1 + p_2 - p_3~. 
\end{align}
In this coordinate system, $\beta$ or $\mu$ can be chosen to be zero due to rotational symmetry around the $z$-axis. A choice of $\mu=0$ leads to
\begin{align}
   p_1 &= 
   (E_1, 0, 0, |{\bm p}_1|)~,  \\
   p_2 &= 
   (E_2, |{\bm p}_2| \sin \alpha \cos \beta~, |{\bm p}_2| \sin \alpha \sin \beta~, |{\bm p}_2| \cos \alpha)~,  \\
   p_3 &= 
   (E_3, |{\bm p}_3| \sin \theta, 0, |{\bm p}_3| \cos \theta)~,  \\
   p_4 &= 
   p_1 + p_2 - p_3~.
\end{align}
From these, the integral over $\mu$ in Eq.~(\ref{eq:collision-term}) can be performed as follows~: 
\begin{align}
   \mathcal{I} = 
   &\frac{1}{4(2\pi)^4} \int \frac{\dd |{\bm p}_2| |{\bm p}_2|^2}{E_2} \frac{\dd |{\bm p}_3| |{\bm p}_3|^2}{E_3} \Lambda(\{f_i\}) \int \dd \cos \theta \dd \cos \alpha \nonumber \\
   &\left. \hspace{18mm} \times \int_0^{2\pi} \dd \beta \delta(p_4^2 - m_4^2) \Theta(p_4^0) \sum_\mathrm{spins} |\mathcal{M}|^2 \right|_{p_4 = p_1 + p_2 - p_3}~.
\end{align}

To use the remaining delta function for the integral over $\beta$, we define 
\begin{align}
   f(\beta)
   &\equiv p_4^2 - m_4^2 \nonumber \\
   &= m_1^2 + m_2^2 + m_3^2 - m_4^2 + 2 p_1 \cdot p_2 - 2 p_2 \cdot p_3 - 2 p_3 \cdot p_1 \nonumber \\
   &= \omega + 2 |{\bm p_2}| \left\{ |{\bm p_3}| \left( \sin \alpha \sin \theta \cos \beta + \cos \alpha \cos \theta \right) - |{\bm p_1}| \cos \alpha \right\}~,
\end{align}
where
\begin{align}
   \omega &\equiv 
   Q + 2 \left(\gamma + |{\bm p_1}| |{\bm p_3}| \cos \theta \right)~,  \\
   Q &\equiv 
   m_1^2 + m_2^2 + m_3^2 - m_4^2~,  \\
   \gamma &\equiv 
   E_1E_2 - E_2E_3 - E_3E_1~.
\end{align}
The integral is transformed using the formula of the delta function
\begin{equation}
   \delta(f(\beta)) = 
   \sum_i \frac{1}{|f'(\beta_i)|}\delta(\beta - \beta_i)~, 
\end{equation}
where $\beta_i$ denotes the solutions of $f(\beta) = 0$.
The condition that $\beta_i$ satisfies is
\begin{equation}
   \cos\beta_i = 
   \frac{1}{2 |{\bm p_2}| |{\bm p_3}| \sin \alpha \sin \theta} \left\{ \omega + 2 |{\bm p_2}| \cos \alpha \left( |{\bm p_1}| - |{\bm p_3}| \cos \theta \right) \right\}~.
\label{cos_beta}
\end{equation}
Thus, there are two solutions $\beta_i = \pm\beta_0~(0 \leq \beta_0 \leq \pi)$.
In this case, $|f'(\beta_i)|$ is expressed as
\begin{align}
   |f'(\beta_i)| 
   &= 
   \left| \mp 2 |{\bm p_2}| |{\bm p_3}| \sin\alpha \sin\theta \sin\beta_0 \right| \nonumber \\
   &= 
   \left|2 |{\bm p_2}| |{\bm p_3}| \sin\alpha \sin\theta \sqrt{1-\cos^2\beta_0} \right| \nonumber \\
   &= 
   \sqrt{(2 |{\bm p_2}| |{\bm p_3}| \sin\alpha \sin\theta)^2 - \left\{ \omega  +2 (|{\bm p_2}| |{\bm p_3}| \cos\alpha \cos\theta - |{\bm p_1}| |{\bm p_2}| \cos\alpha) \right\}^2} \nonumber \\
   &= 
   \sqrt{a \cos^2\alpha + b \cos\alpha + c}~,
\end{align}
where 
\begin{align}
\label{a}
   a &= 
   -4 |{\bm p}_2|^2 (|{\bm p}_1|^2 + |{\bm p}_3|^2 - 2 |{\bm p}_1| |{\bm p}_3| \cos\theta)~, \\
   b &= 
   4 \omega |{\bm p}_2| ( |{\bm p}_1| - |{\bm p}_3| \cos\theta )~,  \\
    c &= 
    4 |{\bm p}_2|^2 |{\bm p}_3|^2 \sin^2\theta - \omega^2~. 
\end{align}
Since the only $\beta$ dependence is of the form $\cos\beta$, both $\beta = \pm\beta_0$ have the same contribution. 
By performing the integration over $\beta$, we obtain
\begin{align}
   \int_0^{2\pi} \dd\beta \delta(f(\beta)) \Theta(p_4^0) \sum_\mathrm{spins} |\mathcal{M}|^2 
   &= 
   \int_{-\pi}^{\pi} \dd\beta \sum_i \frac{1}{|f'(\beta_i)|} \delta(\beta - \beta_i) \Theta(p_4^0) \sum_\mathrm{spins} |\mathcal{M}|^2 \nonumber \\
   &= 
   \frac{2}{|f'(\beta_0)|} \Theta(p_4^0) \sum_\mathrm{spins} \left. |\mathcal{M}|^2 \right|_{\beta = \beta_0} \Theta(\sin^2\beta_0)~. 
\end{align}
The Heaviside step function $\Theta(\sin^2\beta_0)$ must be multiplied because the right-hand side of Eq.~(\ref{cos_beta}) is not guaranteed to take the value in $[-1, 1]$ (a value outside the range leads to $\sin^2\beta_0 < 0$).
Since this Heaviside step function is rewritten as
\begin{align}
   \Theta(\sin^2\beta_0)
   &=
   \Theta ((2 |{\bm p}_2| |{\bm p}_3| \sin\alpha \sin\theta)^2 \sin^2\beta_0) \nonumber \\
   &=
   \Theta(|f'(\beta_0)|^2) \nonumber \\
   &= 
   \Theta(a \cos^2\alpha + b \cos\alpha + c)~,
\end{align}
the final form of the $\beta$ integral becomes
\begin{align}
   \int_0^{2\pi} \dd\beta \delta(f(\beta)) \Theta(p_4^0) \sum_\mathrm{spins} |\mathcal{M}|^2 
   &=
   \frac{2}{\sqrt{a \cos^2\alpha + b \cos\alpha + c}} \Theta(p_4^0) \nonumber \\
   & \qquad \times \sum_\mathrm{spins} \left. |\mathcal{M}|^2 \right|_{\beta = \beta_0} \Theta(a \cos^2\alpha + b \cos\alpha + c)~. 
\end{align}

\bibliographystyle{utphys28mod}
{\small 
\bibliography{ref}
}

\end{document}